\documentclass[a4paper]{jpconf}
\usepackage{graphicx}
\begin{document}
\title{Strange Quark Contribution to the Nucleon Spin from Electroweak Elastic Scattering Data}

\author{S.F. Pate, J.P. Schaub\footnote{Present address:
Department of Physics and Astronomy,
Valparaiso University,
Valparaiso IN 46383, 
USA}}

\address{Physics Department, New Mexico State University, Las Cruces NM 88003, USA}

\ead{pate@nmsu.edu}

\begin{abstract}
The total contribution of strange quarks to the intrinsic spin of the nucleon
can be determined from a measurement of the strange-quark contribution
to the nucleon's elastic axial form factor.
We have studied the strangeness contribution to the elastic
vector and axial form factors of the nucleon, using all available
elastic electroweak scattering data.  
Specifically, we combine elastic $\nu p$ and
$\bar{\nu} p$ scattering cross section data from the Brookhaven
E734 experiment with elastic $ep$ and quasi-elastic $ed$ and
$e$-$^4$He scattering parity-violating asymmetry data from the
SAMPLE, HAPPEx, G0 and PVA4 experiments.  We have not only
determined these form factors at individual values of
momentum-transfer ($Q^2$), as has been done recently, but also
have fit the $Q^2$-dependence of these form factors using simple
functional forms.  We present the results of these fits using
existing data, along with some expectations of how our knowledge
of these form factors can be improved with data from the
MicroBooNE experiment planned at Fermilab.
\end{abstract}

\section{The continuing question of the strange quark contribution to the intrinsic spin of the nucleon}

The techniques of inclusive and semi-inclusive polarized deep-inelastic scattering employed at CERN, SLAC, DESY,
and Jefferson Lab have provided a wealth
of information about the spin structure of the nucleon over the last 25 years.  The contributions of the
$u$ and $d$ quarks in the valence region have now been firmly established.  As well, data from collisions
of polarized protons at RHIC has done much to advance our knowledge of the limitations of the 
gluon spin contribution to the spin of the nucleon.  The strange quark contribution
to the spin of the nucleon, however, is still the subject of investigation, and the indications from deep-inelastic
scattering are unclear at the moment.

Consider the determination of $\Delta s$ + $\Delta \bar{s}$ from {\bf inclusive} 
deep inelastic scattering combined with hyperon $\beta$-decay data; a good example would be the HERMES
measurement~\cite{Airapetian:2007mh} of longitudinal spin asymmetries in 
inclusive positron-proton and positron-deuteron 
deep-inelastic scattering to determine $g_1$ of the proton, deuteron and neutron.  This measurement covered
the kinematic range $0.0041 < x < 0.9$, $0.18~{\rm GeV}^2<Q^2<20~{\rm GeV}^2$, with the data evolved to 
$Q^2$ = 5 GeV$^2$ for analysis.  Using SU(3) flavor symmetry, these data are combined with triplet and octet 
axial charges ($F$ and $D$ from hyperon $\beta$-decay data) in a NNLO analysis to obtain the singlet axial 
charge and the quark contributions to the proton spin.  In doing so, it is necessary to extrapolate the results
into the unmeasured regions of $x$ to fill the interval $0<x<1$.  The result for $\Delta s$ + $\Delta \bar{s}$,
$$\Delta s + \Delta \bar{s} = -0.085 \pm 0.013{\rm (th)}\pm 0.008{\rm (ex)}\pm 0.009{\rm (ev)} 
~~~ {\rm (HERMES~inclusive)}$$
is inconsistent with 0 to more than 4 standard deviations.  

A different technique is used by the same experiment using data 
from {\bf semi-inclusive} deep-inelastic scattering, observing asymmetries in the production of charged pions from 
protons, and in production of charged pions and kaons from deuterons~\cite{Airapetian:2004zf}.
The goal of this analysis is to determine the polarized
parton distribution functions $\Delta q(x)$ (and their integrals)
over the measured $x$-range only; no extrapolations are performed.  As a result, this analysis does not rely on
SU(3) flavor symmetry for combination with triplet and octet axial charges.  However, it is necessary to have
some understanding the of the fragmentation functions which relate the fundamental lepton-quark interaction
to the particles observed in the final state.  The result for $\Delta s(x)$ is consistent with zero in 
the measured $x$-range; the integral of $\Delta s(x)$,
$$\int_{0.023}^{0.30} \Delta s(x) dx = +0.028 \pm 0.033{\rm (stat)}\pm 0.009{\rm (sys)}
~~~ {\rm (HERMES~semi-inclusive)}$$
is then of course consistent with zero as well.

A similar contrasting picture is illustrated by a recent 
global QCD fit by de Florian, Sassot, Stratmann and Vogelsang~\cite{deFlorian:2009vb} 
which is able to bring together
the hyperon $\beta$-decay data, the inclusive and semi-inclusive deep-inelastic data from 
CERN, SLAC, DESY, and Jefferson Lab, and the $\vec{p}\vec{p}$ collision data from RHIC under one roof.  Of special
interest here is their assumption that the strange and anti-strange polarized distributions are equal; 
this is supported by recent COMPASS results~\cite{Alekseev:2010ub}.  Their functional forms allow for a node in the 
polarized parton distribution functions; this permits a small integral of the strangeness polarized
parton distribution function if there is cancelation
of positive and negative contributions from different $x$-regions.  Also, their functional forms allow for SU(2) 
and SU(3) symmetry violation; however, the best fit does not support any
significant deviation from these symmetries.  They determine a ``truncated first moment" of the strange
quark polarized parton distribution function,
$$\int_{0.001}^{1.0} \Delta \bar{s}(x) dx = -0.006^{+0.028}_{-0.031}
~~~ {\rm (DSSV~truncated)}$$
where the uncertainty represents a deviation of the $\chi^2$ from the minimum of the fit by 2\%.
This is clearly consistent with 0.  On the other hand, when the full $x$-range $0<x<1$ is used, the
effect of SU(3) symmetry and hyperon b-decay data is seen:
$$\int_{0}^{1.0} \Delta \bar{s}(x) dx = -0.057
~~~ {\rm (DSSV~full)}$$
which implies $\Delta s + \Delta \bar{s} = -0.114$.  (The authors of 
Ref.~\cite{deFlorian:2009vb} declined to quote an uncertainty for
this result because of the uncertainty in the functions required for extrapolation to $x=0$.)

Additional data, of increased precision, from COMPASS at CERN on {\bf semi-inclusive} deep-inelastic 
scattering~\cite{Alekseev:2010ub}
only deepens the contrast.  Compared to HERMES, this experiment uses a rather different polarized
beam (muons instead of electrons/positrons), polarized target (solid target instead of a gas target), 
and detector system, but the conclusion reached is similar:
$$\int_{0.004}^{0.3} \Delta s(x) dx = -0.01 \pm 0.01{\rm (stat)}\pm 0.01{\rm (sys)}
~~~ {\rm (COMPASS~semi-inclusive)}$$
$\Delta s(x)$ and its integral are consistent with zero in 
the measured $x$-range.\footnote{See also the contribution by Roland Windmolders.} 

\begin{figure}[h]
\includegraphics[width=40pc]{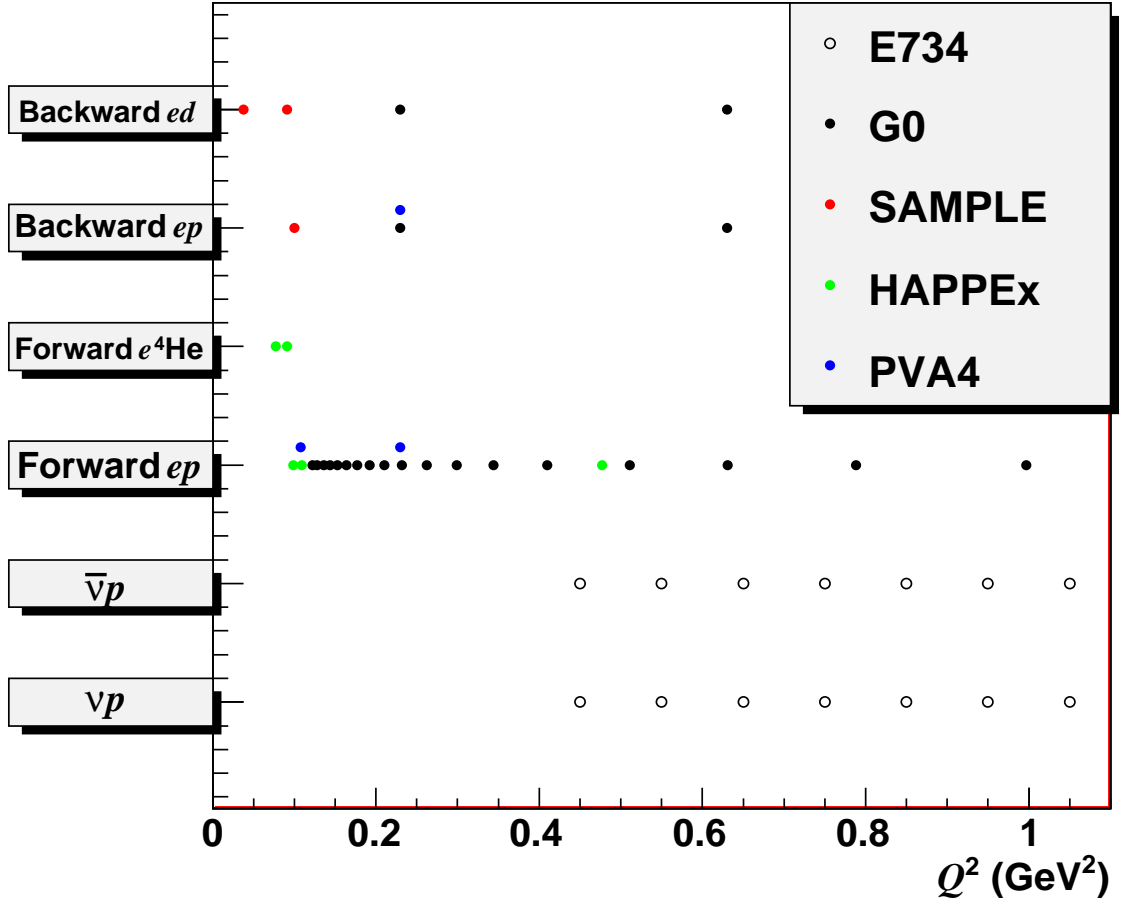}
\caption{\label{datafig} Overview of the types of electroweak elastic scattering data,
and their $Q^2$ values, used in the
analyses for individual values of $Q^2$
summarized in Figure~\ref{fitfig}, as well as in the global fit described herein.
The data are from the BNL E734~\cite{Ahrens:1986xe}, 
G0~\cite{Armstrong:2005hs,Androic:2009zu},
SAMPLE~\cite{Beise:2004py},
HAPPEx~\cite{Aniol:2004hp,Aniol:2005zg,Aniol:2005zf,Acha:2006my}
and PVA4~\cite{Maas:2004ta,Maas:2004dh,Baunack:2009gy} experiments.}
\end{figure}

\section{Strange quark contribution to the vector and axial form factors of the proton}

It is clearly of interest to examine the strange quark contribution to the nucleon spin
in a way that is independent of SU(3) symmetry and 
fragmentation functions.

The full strange quark contribution to the proton spin, $\Delta S$, 
can be directly determined by a measurement of the 
strange contribution to the proton elastic axial form factor, $G_A^s$,
in low energy electroweak elastic scattering.
$$\Delta S \equiv \Delta s + \Delta \bar{s} = G_A^s(Q^2=0)$$
By combining cross sections for $\nu p$ and $\bar{\nu}p$ elastic scattering with parity-violating asymmetries 
observed in $\vec{e}N$ elastic scattering, the strange quark contributions to the nucleon electromagnetic
and axial form factors $G_E^s$, $G_M^s$, and $G_A^s$ may be determined simultaneously~\cite{Pate:2008va,Pate:2003rk}.  
The reasoning behind this method is briefly reviewed here.

In elastic scattering of polarized electrons from unpolarized nucleons, one may observe a parity-violating
helicity-asymmetry in the scattering cross section, $A_L^{PV}$.  Combining this information with the existing data
on the nucleon vector form factors ($G_E^p$, $G_E^n$, $G_M^p$, $G_M^n$), one may extract the strange quark 
contributions $G_E^s$ and $G_M^s$.  It is assumed that charge symmetry is valid, and that the strange quark
distributions in the proton and neutron are identical.  This idea is the basis of the physics programs
of the HAPPEx, G0, SAMPLE, and PVA4 experiments.  An overview of the data from these experiments is given
in Figure~\ref{datafig}.

But the data on $A_L^{PV}$ are largely insensitive to the strange quark contribution to the axial form factor.
To obtain information on $G_A^s$, cross sections for $\nu p$ and $\bar{\nu}p$ elastic scattering are needed;
these cross sections are dominated by the axial form factor at low $Q^2$.  The only existing data for these
cross sections are from the BNL E734 experiment~\cite{Ahrens:1986xe}.

\begin{figure}[h]
\begin{minipage}{18pc}
\includegraphics[trim = 0mm 50mm 0mm 0mm, clip, width=18pc]{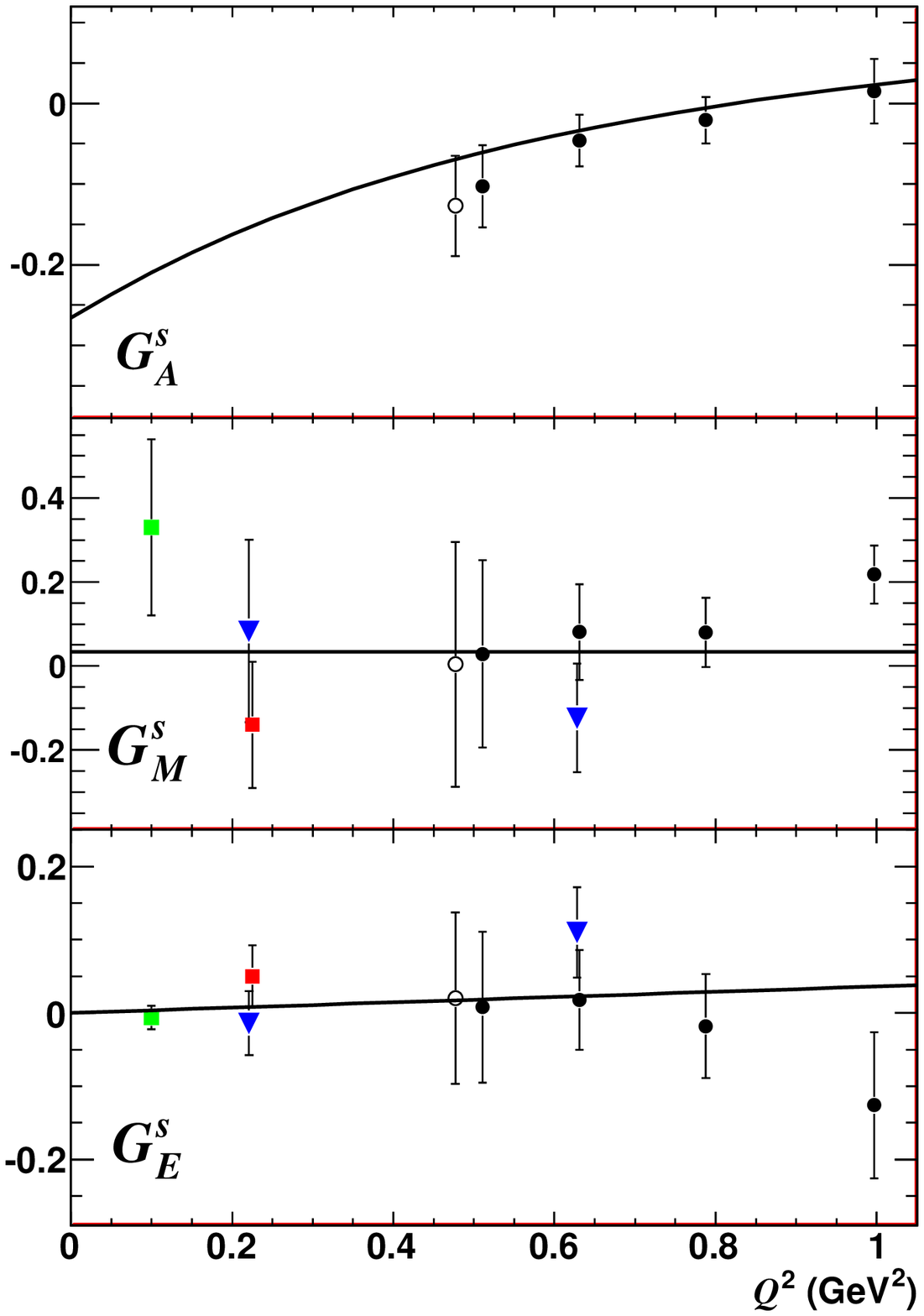}
\end{minipage}\hspace{2pc}%
\begin{minipage}{18pc}
\includegraphics[trim = 0mm 50mm 0mm 0mm, clip, width=18pc]{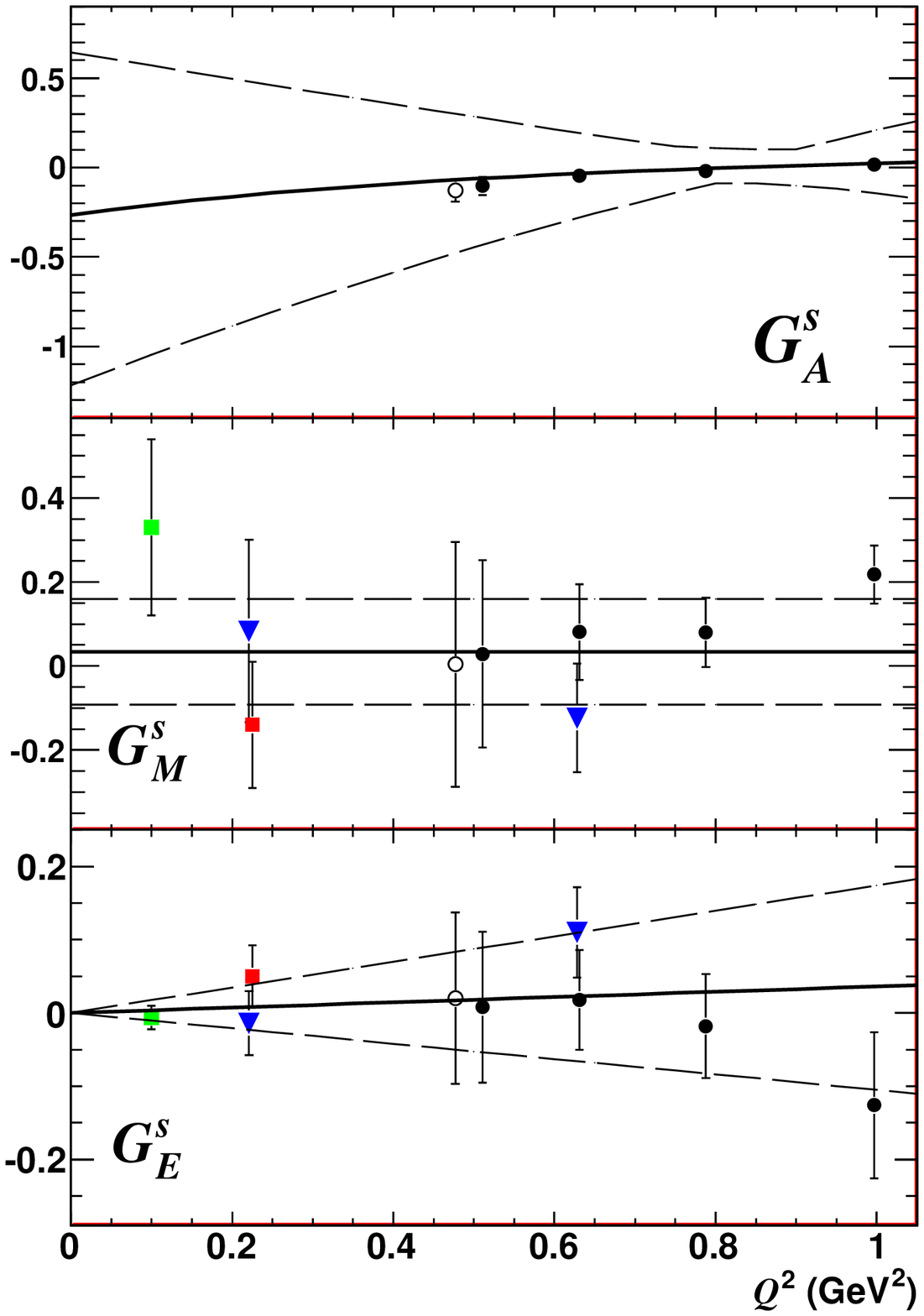}
\end{minipage} 
\caption{\label{fitfig} Results of the determination of $G_E^s$, $G_M^s$, and $G_A^s$ at
individual values of $Q^2$, and also from our global fit.  The separate determinations were done
by Liu et al.~\cite{Liu:2007yi} (green squares at 0.1 GeV$^2$), 
Androi\'{c} et al.~\cite{Androic:2009zu} (blue inverted triangles),
Baunack et al.~\cite{Baunack:2009gy} (red squares at 0.23 GeV$^2$), and
Pate et al.~\cite{Pate:2008va} (open and closed circles).  The preliminary results of the best global fit (see text)
are shown by the solid line; the 70\% confidence level limit curves for the fit are shown
as the dashed line in the right-hand panel.  The vertical scale for $G_A^s$ in the right-hand panel
has been adjusted to accommodate the limit curves of the fit.}
\end{figure}

A number of analyses have already determined $G_E^s$, $G_M^s$ at individual values of $Q^2$, and one
determines $G_A^s$ as well.  These results are reviewed in Figure~\ref{fitfig}.  
Each different analysis has taken some subset
of data covering a narrow range $Q^2$, and uniquely determined the form factors based on standard model expressions
for the asymmetries and cross sections.  The main message from these results is that the strange quark contribution
to the vector form factors is consistent with zero across the full range $0.1~{\rm GeV}^2<Q^2<1.0~{\rm GeV}^2$.  On
the other hand, there is some hint of a signal of a negative $G_A^s$, but a lack of neutrino-scattering data
at low $Q^2$ prevents a definite conclusion about $\Delta S \equiv \Delta s + \Delta \bar{s} = G_A^s(Q^2=0)$ 
at this time.

\section{Global fit of electroweak elastic scattering data}

We have performed a global fit of the available electroweak elastic scattering data, using the same general 
technique described in Ref.~\cite{Pate:2008va}, but now including all of the data
in Figure~\ref{datafig} and assuming functional forms for the form factors.
A fit of this type can 
quantity the amount of information that can be extracted from the electroweak data in 
Figure~\ref{datafig}, provide a mechanism for including new data that may become available,
and can be used as a tool to estimate the impact of any future experiments seeking to improve
our knowledge of the strangeness form factors.
  
Our first attempt to fit the form factors $G_E^s$, $G_M^s$, and $G_A^s$ used this 
simple set of functional forms:
$$G_E^s = \frac{\rho_s\tau}{(1+Q^2/\Lambda_E^2)^2} ~~~~ G_M^s = \frac{\mu_s}{(1+Q^2/\Lambda_M^2)^2}
~~~~ G_A^s = \frac{\Delta S}{(1+Q^2/\Lambda_A^2)^2}~~~~{\rm (initial~function~set)}$$
where $\tau = Q^2/4M_N^2$, $\rho_s \equiv (dG_E^s/d\tau)|_{\tau=0}$ is the strangeness radius, $\mu_s$ is
the strangeness magnetic moment, and the $\Lambda$s are shape parameters that define how rapidly these functions
depend on $Q^2$.  These functions satisfy a number of boundary conditions: they 
all depend only on even powers of $Q$; they tend towards 0 as $Q^2\rightarrow\infty$, and they come to the correct
physical value at $Q^2=0$.  We found, however, that this set of functions is not quite suited to the existing data.
First of all, a good fit to the neutrino data was not possible unless very large negative 
(and clearly unphysical) values of $\Delta S$
were permitted; this problem was solved by introducing a term linear in $Q^2$ in the numerator of the
$G_A^s$ function.  Secondly, the shape parameters $\Lambda_E$ and $\Lambda_M$ are not determined; the best fit
values for them are very large, with very large errors.  This is not surprising; looking at Figure~\ref{fitfig}
it seems that both $G_E^s$ and $G_M^s$ are essentially featureless, since they are consistent with zero.  
If $\Lambda_E$ and $\Lambda_M$ are very large, then the $Q^2$-dependence in the denominators of $G_E^s$ and
$G_M^s$ may be neglected and we may use a simple linear function for $G_E^s$ and a constant function for $G_M^s$.
Then our final set of functions were these.
$$G_E^s = \rho_s\tau ~~~~~~ G_M^s = \mu_s
~~~~~~ G_A^s = \frac{\Delta S + S_A Q^2}{(1+Q^2/\Lambda_A^2)^2}~~~{\rm (final~function~set)}$$
The best values for the five parameters (preliminary) are:
$$ \rho_s = 0.13 \pm 0.21 ~~~ \mu_s = 0.035 \pm 0.053 ~~~ \Delta S = -0.27 \pm 0.41 ~~~ \Lambda_A = 1.3 \pm 1.9
~~~ S_A = 0.32 \pm 0.48$$
The best fit is shown as the solid line in Figure~\ref{fitfig}; the 70\% confidence level uncertainty limits
are shown by the dashed line.

It should be stressed that the fit has {\bf not} been made to the values of the form factors displayed in 
Figure~\ref{fitfig}.  Instead, the fit has been made to the measured values of the cross sections and asymmetries
from the experiments mentioned in Figure~\ref{datafig}.  In Figure~\ref{fitfig}, we show 
that this global fit is consistent
with the results of previous limited analyses of subsets of these data.

The preliminary results of the fit can be simply described.  The strangeness radius and magnetic moment are consistent with zero, and the uncertainties in these parameters are consistent with the uncertainties
in the separate determinations of $G_E^s$ and $G_M^s$ on display in Figure~\ref{fitfig}.  On the other hand,
$\Delta S$ is also consistent with 0 but the uncertainty is very large because there are no $\nu p$ or
$\bar{\nu} p$ elastic data at sufficiently low $Q^2$ to constrain it.  As a result the uncertainties in the
global fit to $G_A^s$ are very much larger than the uncertainties
in the separate determinations of $G_A^s$ in Figure~\ref{fitfig}.

It is clear that the single greatest need is for a much improved set of data on neutral current
$\nu p$ and $\bar{\nu} p$ elastic scattering down to the lowest $Q^2$ possible.  
In the last few years it has become clear that 
a better understanding of the charged-current cross section is also needed, as described in the next section.

\section{The problem of the charged-current axial form factor}

The results of this fit are still preliminary because we do not have a full systematic uncertainty analysis.  Such
an analysis would consider variations in all the fixed input parameters, such as the 
full nucleon form factors, within
their known limits as determined by existing experimental data or by theoretical considerations.

One nucleon form factor that must be known for this analysis to proceed is the charged-current 
portion of the axial form factor.  The full nucleon axial form factor contains contributions from
up, down, and strange quarks,
$$G_A^Z = \frac{1}{2}\left(-G_A^u + G_A^d + G_A^s\right)$$
and this form factor occurs in the current associated with $Z$-exchange processes like
$ep$, $\nu p$, and $\bar{\nu}p$ elastic scattering.
The ``charged-current" portion of that form factor,
$$G_A^{CC} = G_A^u-G_A^d$$
occurs in $W$-exchange processes, like $\nu_\mu + n \rightarrow \mu^- + p$.  The cross section for
$\nu_\mu + n \rightarrow \mu^- + p$ has been studied using deuterium and nuclear targets for many
years, and until a few years ago the form factor $G_A^{CC}$ extracted from those cross sections was
considered ``known."  In this analysis we used the parametrization provided by 
Budd, Bodek and Arrington~\cite{Budd:2003wb,Bodek:2003ed,Budd:2004bp}.  In the analysis technique
used here and in Ref.~\cite{Pate:2008va}, it is the full axial form factor $G_A^Z$ which is determined from
the neutral current $ep$, $\nu p$, and $\bar{\nu}p$ data; the charged-current part $G_A^{CC}$ must be subtacted
away to obtain $G_A^s$; any uncertainty in $G_A^{CC}$ translates directly into an uncertainty in $G_A^s$.

In recent years, the charged-current portion of the axial form factor has become a lot less
``known" than it used to be, due to high-statistics neutrino-nucleus scattering experiments such as
MiniBooNE~\cite{AguilarArevalo:2007ru,AguilarArevalo:2010zc} and K2K~\cite{Gran:2006jn} that raise
questions about our knowledge of its $Q^2$-dependence.  
Due to the imperfect knowledge of the incoming neutrino momentum and inability to reconstruct the vertex 
kinematics, a more complete and sophisticated way of analysing these data must be developed which
takes into account the electroweak nuclear response.  
This is underway -- see for example Benhar, Coletti, and Meloni~\cite{Benhar:2010nx},
Ankowski, Behnar, and Farina~\cite{Ankowski:2010yh}, and references therein.

In order to improve the situation, it is necessary 
to have both the charged-current and neutral-current elastic channels measured at low $Q^2$ in the same 
experiment and treated in a consistent way to extract the axial form factor(s).  
The MicroBooNE experiment at Fermilab may provide these data.

MicroBooNE is an approved experiment at Fermilab to build a large liquid Argon Time Projection Chamber (LArTPC) 
to be exposed to the Booster neutrino beam and the NuMI beam at Fermilab. The experiment will 
address the low energy excess observed by the MiniBooNE experiment, measure low energy neutrino 
cross sections, and serve as the necessary next step in a phased program towards massive Liquid Argon TPC detectors.

Liquid argon is very well-suited to observe low energy protons from low-$Q^2$ neutral-current and 
charged-current scattering; this experiment may be an ideal place to move towards a successful 
determination of $\Delta s$.

\section{Summary}

The strangeness contribution to the nucleon spin may be determined from a combined analysis of 
low-energy electron-nucleon and neutrino-nucleon elastic scattering data.

However, existing neutral-current neutrino data (from BNL E734) lack insufficient precision and 
$Q^2$-range to make possible a definitive determination.  In addition, recent neutrino 
experiments have called into question our understanding of the charged-current axial form factor 
extracted from data on nuclear targets.

New experiments (e.g. MicroBooNE) can provide the datasets needed for a consistent 
treatment of the electroweak nuclear response and extraction of the strange axial form factor.

\ack{This work was funded by the US Department of Energy, Office of Science.}

\section*{References}
\bibliography{SPIN2010_Pate}

\end{document}